\documentclass[prb,showpacs,twocolumn]{revtex4-1}
\usepackage{amssymb}
\usepackage{graphicx}
\usepackage{bm}
\usepackage{xcolor}
\usepackage{epsfig}
\usepackage{amsfonts}

\begin{document}
\title{Mesoscopic fluctuations in  biharmonically driven flux qubits}
\author{Alejandro Ferr\'on}
\affiliation{Instituto de Modelado e Innovaci\'on Tecnol\'ogica (CONICET-UNNE)and Facultad de Ciencias Exactas, Naturales y Agrimensura, Universidad Nacional del Nordeste, Avenida Libertad 5400, W3404AAS Corrientes, Argentina.}
\author{ Daniel Dom\'{\i}nguez and Mar\'{\i}a Jos\'{e} S\'{a}nchez }
\affiliation{Centro At{\'{o}}mico Bariloche and Instituto Balseiro, 8400 San Carlos de Bariloche, Argentina.}

\begin{abstract}

%Weakly disordered electronic mesoscopic systems exhibit at milikelvin temperatures fingerprints of phase coherence. Among others phenomena, weak localization and universal conductance fluctuations have been studied for almost thirty years.
 %More recently, signatures of coherence have been also analized  in the dynamics of superconducting solid state devices controlled by  external (dc + ac)  magnetic fields. 
We  investigate flux qubits
 driven by a biharmonic  magnetic  signal,  with a phase lag that acts as an effective  time reversal broken parameter. 
 The driving induced transition rate   between the ground and  the excited state of the flux  qubit can be thought as  an effective transmitance, 
profiting from  a direct analogy between interference effects at   avoided level crossings and scattering events in  disordered  electronic systems.
For time scales prior to full relaxation but large compared to the decoherence time, this characteristic rate  has been accessed experimentally and its sensitivity 
 with  both the phase lag and the dc flux detuning explored. In this way signatures of Universal Conductance Fluctuations-like effects have recently been analized  in flux qubits and compared with a  phenomenological model that only  accounts  for decoherence, as a classical noise.
We here  solve the full  dynamics of the driven flux qubit  in contact with a quantum  bath  employing  the Floquet Markov Master equation. Within this formalism relaxation and decoherence rates result strongly dependent on both the phase lag and 
the dc flux detuning.
% with values that can differ  in an  order of magnitude with  those predicted for the  undriving system and reported experimentally. Morover, close  to resonances with the driving field, decoherence and relaxation rate become
%comparable. 
Consequently, the associated pattern of fluctuations in the characteristic rates   display important differences with those   obtained  within the mentioned phenomenological model.
In particular we demonstrate  the  Weak Localization-like effect in the averages values of the relaxation rate. 
Our  predictions    can   be tested for  accessible, but longer  time scales than the  current experimental times.

\end{abstract}

%\pacs{74.50.+r,85.25.Cp,03.67.Lx,42.50.Hz}

\maketitle

\section{Introduction}
\label{int}

The quantum conductance of a phase coherent conductor can be related, in the diffusive regime, 
to the transmission probability through the disordered  region.\cite{datta}
 For milikelvin temperatures,  when typically the coherence length could become  larger than the scattering mean free path, the interference term present
in the transmission probability survives disordered averaging, giving  rise to quantum corrections to the classical transport properties and novel phenomena.
 Mesoscopic effects like Weak Localization   and Universal Conductance Fluctuations
have been predicted and extensively tested in electronic quantum system for years.\cite{varios_ucf,varios_wl,leshouches}

Universal conductance fluctuations (UCF) are  sample to sample fluctuations- of the order of the  quantum of  conductance- originated on  the  sensitivity of the quantum conductance 
to changes in  an external parameter, like a magnetic flux or a gate voltage.\cite{varios_ucf}

The Weak localization (WL) effect is a quantum correction to 
the classical conductance, that survives disorder averaging when time reversal symmetry is present.\cite{varios_wl}
Without spin-orbit effects, it is characterized by   a dip in the conductance (peak in the resistance) at zero magnetic field. The standard way to detect the WL  effect is by its suppression, as its strength falls off with  an applied magnetic field. A critical  field that scales as $B_{c} \sim 1/ ( D \tau_{\phi})$, with $D$ the diffusion coefficient and  $\tau_{\phi}$ the coherence time, washes out  
the quantum interference term, and thus the WL  correction. \cite{varios_wl}
The measurement of $B_{c}$  has been established as a  usual route to determine the coherence time. \cite{varios_wl,leshouches}

Flux qubits (FQ)  are model artificial atoms whose  energy levels  can be manipulated by an external magnetic flux. \cite{qbit_mooij,chiorescu} For most of the applications in 
 quantum information theory, only the  two lowest energy levels  of the FQ have been considered in  studies of their quantum dynamics.\cite{chiorescu}
 However, FQ  exhibits as a function of the static magnetic flux a complex structure  of energy levels with multiple  avoided crossings. This rich  spectrum   can be explored  by driving the FQ  with an ac  magnetic flux, for moderate driving frequencies - in the microwave range.
In a typical protocol, the FQ is initially prepared  in its  ground state for a given value of the dc flux, and evolves under the ac driving  quasi adiabatically until the first avoided crossing is reached. There, the  state obeys a  Landau-Zener-St\"uckelberg (LZS)  transition  and transforms  into a coherent superposition of the ground and excited states.\cite{oliver,nori2011} 
For weak  ac  amplitudes, such that a single avoided crossing is reached by the driving protocol, the superposition state and the initial one interfere again at the second passage for the avoided crossing. Hence, the avoided crossing acts as an effective beam splitter, where scattering events take place.
For driving periods larger than the coherence time, the evolved state  accumulates 
a total phase, that depends  both on the dc flux and the driving amplitude.\cite{oliver,nori2011}.

The regime of weak driving, when  only the lowest two energy levels of the FQ are explored
and a single avoided crossing is attained by the amplitude of the ac flux,
has been studied in extent both experimentally and theoretically. \cite{oliver,ferron,gustavsson}
 In this way, FQ  have been investigated extensively in recent years as high resolution Mach-Zhender type of interferometers. \cite{oliver}

%For large driving amplitudes, when  many avoided crossings can be reached, the repeated sweeps 
%through the avoided level crossings result in successive LZS  transitions
% between different energy levels. After many driving periods - and due to the sensitivity of the total  accumulated phase to the dc flux and the driving amplitude- the FQ population displays a rich pattern of interference fringes, reminiscent to those generated 
%by coherent light  scattered 
%in a disordered potential \cite{cbl}. 
%This driving  protocol -named  
%as amplitude spectroscopy (AS) \cite{oliver}- was employed

For large driving amplitudes, when  many avoided crossings can be reached, the repeated sweeps 
through the avoided level crossings result in successive LZS  transitions
between different energy levels.
This driving  protocol -named  as amplitude spectroscopy (AS) \cite{oliver}- was employed
to reconstruct the  FQ energy level spectrum and to study its dynamics under different conditions. \cite{valenzuela,fds}  The AS protocol has  been also successfully applied in other  systems like 
charge qubits \cite{cqubits}, ultracold molecular gases \cite{mark} and single electron spins. \cite{huang}

The interaction  of the driven  FQ with an external bath has been recently studied  to  incorporate  more realistic dissipative scenarios beyond the pure coherent regime.
Relevant and potential useful  phenomena like population inversion,  \cite{sun2009,hanggi,ferron_prl}   dynamical transition in the interference patterns, \cite{ferron_arx} and  estimates for coherence times have been extracted from these studies.\cite{shevchenko, dupont2013}

While  a priori there is not a direct connection between driven FQ and  mesoscopic    disordered
electronic systems, the identification of a transition at the avoided crossing as  a   scattering  event, suggests a route  to study mesoscopic-like effects in driven FQ.
However, for the weak driving  regime at most one avoided crossing is reached by the protocol and only two scattering events (transitions) take place in one period of the driving. This poor scattering regime seems  insufficient to explore the mesoscopic analogy.

An alternative %to go beyond this limitation but keeping the weak driving regime,%
 was proposed in Ref.\onlinecite{gustavsson} with the implementation of a protocol generated by a biharmonic flux with a phase lag.
The signal was designed to drive the FQ  up to four times through the avoided crossing in one period, which was chosen much shorter than the  energy relaxation time. 
Therefore, after many periods of the driving the excited state of the FQ was populated as a function of time with  a  characteristic  (equilibration) rate that was extracted in the experiment by a fitting procedure.

Other interesting   phenomena of biharmonic drive in flux qubits have been
studied in recent experimental \cite{bylan2009, forster2015} and theoretical \cite{satanin2014,blattmann2015} works.

The interference conditions  can be changed by either tuning the external dc flux or the phase lag in the driving wave form.  Following this strategy,  the equilibration rate $\Gamma$ and its concomitant  fluctuations have been  analyzed in Ref. \onlinecite{gustavsson}.

For large driving frequencies and for time scales smaller than the relaxation time but larger than the decoherence time, it is possible to study  the dynamics of the FQ within a model of classical diagonal noise and computing   $\Gamma$ from phenomenological rate equations.\cite{lt} 
Neglecting relaxation, it can be shown that $\Gamma \sim 2 W$, with $W$  the transition rate induced by driving. \cite{lt} 
The mesoscopic analogy proposed in  Ref.\onlinecite{gustavsson} was to identify  $W$    with  a  transmission  rate, which in (mesoscopic) electronic transport determines   the conductance. \cite{datta} Thus the  goal was  to access  the fluctuations in $W$  through the study of the fluctuations in  $\Gamma$.
This scenario, although tempting, should be taken with caution.

As we already mentioned, the expression $\Gamma \sim 2 W$ is valid for large driving frequencies and for time scales far below relaxation. \cite{berns,gustavsson}
%However, we will show that important discrepancies  emerge with the values of $\Gamma$ 
%obtained from the stationary state, after complete relaxation.
 %and those computed for  the experimental time scales -which are far below the asymptotic regime.\cite{oliver}

We solve  the  full dynamics of the driven FQ  employing  the Floquet Markov (FM) master  equation to include  relaxation and decoherence processes for a realistic model of a quantum bath.\cite{hanggi, ferron_prl, ferron_arx}
This   formalism \cite{floquet},  valid for  arbitrary time scales and strength of the driving protocol, allows  to compute the decoherence and the relaxation (equilibration) rates. 
As we will show, both rates  result strongly dependent on the driving amplitude and the dc flux, attaining values that might differ up to  an order of magnitude from those determined in the absence of driving.
%Additionally important discrepancies  will emerge between  the values of the relaxation rate
%obtained from the stationary state, and the one computed for  experimental time scales -which as we show are far below the asymptotic regime.
Consequently, the  relaxation (equilibration) rate  obtained within the FM formulation might strongly differ from   the value $2 W$ - used in  Ref.\onlinecite{gustavsson}  to compare with  the experimental results-.

%On the other hand,  the reported  values for the driving period $\tau$ and the  decoherence time,$T_{2}$, satisfies $\tau \lesssim T_{2}<< T_{1}$.\cite{gustavsson} These time scales put the experiment a priori in a quasi classical regime, that preserves coherence  within one period of the driving  $\tau$  without resolving  multiphoton resonances.\cite{berns}
%However,  will show that relaxation and decoherence rates are  strongly modified by the driving parameters.  \cite{ferron_arx}.
% A similar behavior should be expected for the biharmonic driving protocol implemented in  Ref.\onlinecite{gustavsson}. 

%Due to the implementation o the biharmonic driving, the number of scattering events per  period was increased from two up to four. However the  experiment still remains in the few scatterers limit.
%One of the goals of the present  work is to go beyond this limit in order to reach the truly ``diffusive" regime and
%to explore  new features  that emerge  when the FQ is strongly driven.

An important outcome of our study is related to the weak localization (WL) effect, which was not resolved in the experiment of Ref.\onlinecite{gustavsson}.
As we will analize, %the WL correction could be detected in the regime of larger coherence times, $T_{2} > \tau $. It% 
it  is not the driving protocol but   the accessible decoherence time which limits the  detection of the effect.  \cite{gustavsson}. In fact, the WL correction could be measured in the regime of larger coherence times. 

The paper is organized as follows.  In Sec. \ref{fq} we review the  Hamiltonian model for the flux qubit (FQ) and the effective Hamiltonian obtained when only the two lowest  levels of the FQ are
considered. 
In Sec.\ref{stls} we  
 derive  an analytical  expression 
for the    rate  $\Gamma$ obtained within a phenomenological approach which includes  classical noise  as the only  source of decoherence. Gaussian  and  low frequency  type of noise are both considered. \cite{paladino}.

Due to the limitations of the analytical approach, already mentioned,  
we implement in Sec.\ref{fm} the full quantum mechanical calculation in order   to obtain the equilibration (relaxation)  rate $\Gamma_r$  within the Floquet Markov (FM) master equation.
The last part of this section is devoted  to compare  
the behavior of $\Gamma \sim 2 W $ and $\Gamma_r$  as a function of the dc flux, and to analize the effect that the driving  has on the determination of both decoherence and relaxation rates.
The  fluctuations in the  rates   $\Gamma \sim 2 W $ and $\Gamma_r$ as a function of the dc flux and the time reversal parameter are analized in Sec.\ref{fluct}.
As we show, besides UCF,  clear signatures of WL correction could  be also detected if the coherence
 is increased.
We discuss  the limitation imposed by 
the accessible decoherence time in the experimental determination of  the WL correction. 
We conclude in Sec.\ref{conc} with a  discussion and perspectives.

\section{The flux qubit}\label{fq}

The Flux qubit (FQ) consists on a superconducting ring with three Josephson
junctions\cite{qbit_mooij} enclosing a magnetic flux $\Phi=f\Phi_0$, 
with $\Phi_0=h/2e$. 
Two of the junctions have the same Josephson coupling energy $E_{J,1}=E_{J,2}=E_J$, and 
capacitance, $C_1=C_2=C$, while the third one has smaller coupling 
$E_{J,3}=\alpha E_J$ and capacitance $C_3=\delta C$, with $0.5<\delta<1$.
The junctions have gauge invariant phase differences defined as 
$\varphi_1$, $\varphi_2$ and $\varphi_3$, respectively. Typically the circuit 
inductance can be neglected  and the phase difference of the third junction 
is: $\varphi_3=-\varphi _1 +\varphi _2 -2\pi f$.

 %For millikelvin temperatures, it behaves as an artificial atom with
 %quantized energies levels that are sensitive to an external magnetic 
%field. \cite{qbit_mooij,chiorescu,revqubits}

Therefore, the system can be 
described in terms of two independent dynamical variables, chosen as 
$\varphi_l=(\varphi_1-\varphi_2)/2$ (longitudinal phase) and
$\varphi_t=(\varphi_1+\varphi_2)/2$ (transverse phase).  In terms of these 
variables, the hamiltonian of the FQ (in units of $E_J$) is: \cite{qbit_mooij}

\begin{equation}
	\label{ham-sys}
	{\cal H}_{FQ}=-\frac{\eta^2}{4}\left(\frac{\partial^2}{\partial\varphi_t^2}+
	\frac{1}{1+2\delta}\frac{\partial^2}{\partial\varphi_l^2}\right)
	+V(\varphi_l,\varphi_t)\; ,
\end{equation}
with $\eta^2=8E_C/E_J$ and $E_C=e^2/2C$. The kinetic term  
corresponds to the electrostatic energy of the system  and the potential one to the 
Josephson energy of the junctions, given by
$V(\varphi_l,\varphi_t)= 2+\delta -2\cos\varphi_t\cos \varphi_l - 
\delta \cos (2\pi f+2\varphi _l)$.
%\begin{equation}
%\label{eq:pot}
%V(\varphi_l,\varphi_t)= 2+\alpha -
%2\cos\varphi_t\cos \varphi_l - \alpha \cos (2\pi f+2\varphi _l) \;
%\end{equation}
Typical FQ  experiments  have values of $\delta$ in the
range $0.6-0.9$ and $\eta$ in the range $0.1-0.6$. \cite{chiorescu,valenzuela}
It is operated at  magnetic fields near the half-flux quantum, \cite{qbit_mooij,chiorescu}  
$f=1/2+f_0$, with $f_0 \ll 1$. For $\delta \ge 1/2$, the potential 
$V(\varphi_l,\varphi_t)$ has two minima at 
$(\varphi_l,\varphi_t)=(\pm\varphi^*,0)$ separated by a maximum at 
$(\varphi_l,\varphi_t)=(0,0)$. Each minima corresponds to macroscopic 
persistent currents of opposite sign, and for $f\gtrsim 1/2$ 
($ f \lesssim 1/2$) a ground state $|+\rangle$ ($|-\rangle$) with positive 
(negative) loop current is favoured.

For values of  $|f_0| \ll 1$, such that the avoided crossings with the third energy 
level are not reached, the hamiltonian  of Eq.~(\ref{ham-sys}) 
can be reduced to the two-level system (TLS)\cite{qbit_mooij,ferron} 

%studied in Sec. \ref{stls}:

\begin{equation}\label{htls}
 {\cal H}=-\frac{\varepsilon_0}{2} {\hat\sigma}_z - \frac{\Delta}{2} {\hat\sigma}_x
\;,
\end{equation}
in the basis defined by the persistent current states 
$|\pm\rangle=(|0\rangle\pm|1\rangle)/\sqrt{2}$, with $\hat{\sigma}_z$, $\hat{\sigma}_x$ the Pauli matrices and $|0\rangle$ and $|1\rangle$ the ground and excited states at $f_0=0$. The parameters of 
${\cal H}$ are the gap (at $f_0 =0$) $\Delta$ and the detuning 
energy $\varepsilon_0= 4\pi I_p  f_0$. Here 
$I_p=\delta |\langle+|\sin2\varphi _l|+\rangle|=\delta |\langle-|\sin2
\varphi _l|-\rangle|$ is the magnitude of the loop current, which for our case 
with $\delta=0.8$ and $\eta=0.25$  is $I_p=0.721$ (in units of 
$I_c=2\pi E_J/\Phi_0$).

\begin{figure}[h]
\begin{center}
\includegraphics[width=1.\linewidth,clip]{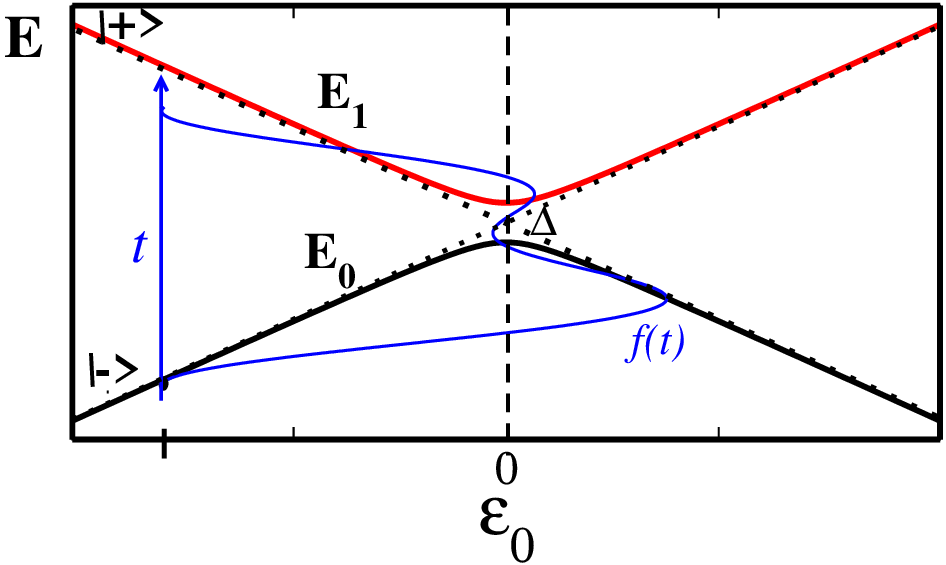}
\end{center}
\caption{(color online) Flux Qubit energy diagram for the TLS hamiltonian
 Eq.(\ref{htls}),  as a function of flux detuning
$\varepsilon_{0}$. Inset:   implemented biharmonic pulse  $f(t)$, chosen to drive  the FQ  
four times through  the avoided crossing in one period.}
%(b) Simulated transition rates $W$ using modified
%Eq.\ref{eqb} for qubit parameters $\Delta/h=19\,MHz$, $\omega_0/(2\pi)=
%125\,MHz$, $A_{1}=3\,m\Phi_0$ and $A_{2}=1.65\, m\Phi_0$,
%$\alpha=0$ and different
%values of $T_2$, $\Gamma_2=200\,MHz$ (black line) and $\Gamma_2=20\,MHz$ (red &line).
\label{f1}
\end{figure}

Figure \ref{f1} sketches the energy levels diagram for the FQ hamiltonian restricted to the TLS,
 Eq.(\ref{htls}) with $E_{0,1}= \pm 1/ 2 \sqrt {\varepsilon_0^2 +  \Delta^2}$
the ground and excited states energies, respectively.
\cite{oliver,valenzuela,berns}.

The FQ restricted to weak driving amplitudes was the regime explored in Ref.\onlinecite{gustavsson}. Consistently, in the following we  focus on the dynamics of the TLS Hamiltonian  Eq.(\ref{htls}) under the effect of the  biharmonic driving.

\section{Two level system under biharmonic driving}

\subsection{Equilibration rate within the classical noise model} 
\label{stls}

%We analize the dynamics of the  FQ restricted to the TLS, disregarding additional levels not explored %	so far in the experiment of Ref.\onlinecite{gustavsson}.

As we mentioned, in Ref.\onlinecite{gustavsson} the equilibration rate  $\Gamma$ is experimentally  determined by  fitting the  decay of the excited population to the equilibrium, assuming  an exponential behavior as a function of time. 

In the following we describe a route to compute $\Gamma$ from  phenomenological rate equations  in the regime of  large driving frequencies, for  which the change in the qubit population (per unit time) induced by the driving  is  small compared to the decoherence  
rate $\Gamma_2 \equiv 1/T_2$  but large compared to the inelastic relaxation rate  in the absence of driving, $\Gamma_1\equiv 1/T_1$. As the calculation is adapted from the one derived for the  single driving protocol \cite{berns} we 
here present the main steps stressing  differences.
The source of noise is considered classical and diagonal, which essentially means that the noise  
produces pure dephasing.  However diagonal noise is  consistent with  the typical  experimental situation  with FQ where the dominant source of  noise is flux-like.

%In the next section  we will develop
%the full quantum treatment employing the Floquet Markov approach, valid 
%even  for times scales of the order of $T_1$.

From phenomenological rate equations %and for times scales $t$ such  that  $ T_{2} \ll t < T_{1}$,%
 the equilibration rate can be written as   
$\Gamma= 2 W +  \Gamma_1$ , with  $W$ the transition rate induced by the driving protocol.\cite{berns} 
For large relaxation times $T_1$ and for  $W \gg \Gamma_{1}$, one gets $\Gamma \approxeq 2 W$. 
Larger  values of $\Gamma_{1}$ would require the explicit inclusion of  
relaxation processes in the analysis  to avoid important differences between $W$ and  $\Gamma$. 
These cases will be addressed in Sec.\ref{fm}.

To compute $W$, we include  in Eq.(\ref{htls}) the time dependent biharmonic driving  and  the diagonal classical  noise
by  replacing  $\varepsilon_0 \rightarrow h(t)= \varepsilon_0 + \delta\varepsilon + \varepsilon(t)$.
The term  $\varepsilon(t)=  4 \pi I_p f(t)$, contains the biharmonic ac flux 
 $f(t)= A_{1}\cos{(\omega_0 t+\alpha)}-A_{2}\cos{(2\omega_0 t)}$ of fundamental frequency $\omega_0= 2 \pi / \tau$.  The phase lag $\alpha$ turns the protocol asymmetric  in time and the amplitudes ratio  $A_{1}/A_{2}$ was chosen to drive the FQ  up to four times through the avoided crossing in one period $\tau$. The  classical  noise is   $\delta\varepsilon$. 

 %The driving
 %waveform $f(t)$ was chosen to traverse the avoided crossing four times in one period $\tau$.

%To obtain  $W$, we adapt to the biharmonic protocol, the steps followed  in  
%Ref. \onlinecite{berns} for the single driving case.

The Hamiltonian Eq.(\ref{htls}) can be  turned  to purely off-diagonal becoming, after an unitary transformation (we use $\hbar = 1$), 
%$\exp{i \phi(t) \hat{sigma}_z}$ with $\phi (t)=\int_{0}^{t} h(t^') dt^'$, 
%that turns the hamiltonian   purely off-diagonal,

\[
\tilde{{\cal H}}=-\frac{\Delta^\ast(t)}{2}\hat{\sigma}_x \; ,
\]
\noindent where  $\Delta(t)=\Delta e^{-i\phi(t)}$ and $\phi(t)=\int_{0}^{t} h(t') dt'$.

The FQ is usually initially prepared in its ground state, $\vert \Psi_g (t=0)\rangle $,
for a given value of the detuning $\varepsilon_0$
(in Fig.\ref{f1} an initial state for $\varepsilon_0  < 0$ is chosen). 
Alternatively it is possible 
to initialize the FQ in an eigenstate of the computational basis, {\it i.e.} $\vert -\rangle$($\vert +\rangle$)
 for $\varepsilon_0  < 0$ ($\varepsilon_0  > 0$).
In general, for values of flux detuning  
$\varepsilon_0 > \Delta $, the initial state satisfies
$\vert \Psi_g (t=0)\rangle \longrightarrow \vert -\rangle (\vert +\rangle)$ for 
$\varepsilon_0 < 0 (\varepsilon_0 > 0)$.

The transition rate $W$ induced by the driving is the time derivative of the transition probability between the initial and the final state. Under the assumption of fast driving,  $\omega_0= 2 {\pi/\tau} > \Delta$,
 it can be computed expanding the time evolution operator  
to first order in $\Delta$,
\[
U(t, 0)= 1 -i \int_{0}^t \tilde{{\cal H}} (\tau) d \tau + {\cal O} ( {\Delta ^2}) \; . 
\] 

Therefore we write:

%\begin{equation}
%W=\frac{d}{dt} \vert \langle 1 \vert  U(t) \vert 0\rangle\vert^ 2\lim_{\delta t\rightarrow\infty}\frac{|A_{t,t'}|^2}{\delta t}\;\;\;\;
%A_{t,t'}=\frac{1}{2}\int_{t'}^t \Delta(\tau)d\tau \; .
%\end{equation}

\begin{eqnarray}\label{rate}
&&W= \frac{d}{dt} \vert \langle + \vert  U(t,0) \vert -\rangle\vert^ 2 = \frac{d}{dt} \frac{1}{4}\int_{0}^t \int_{0}^t \Delta(\tau_{1}) \Delta^\ast(\tau_{2})  d\tau_{1} d\tau_{2} \;     \nonumber \\ 
&& = \frac{d}{dt} \frac{\Delta^2}{4} \int_{0}^t \int_{0}^t {\Re}\lbrace \exp-i \left( \phi(\tau_1)-\phi(\tau_2)\right) \rbrace d\tau_{1} d\tau_{2}  , 
\end{eqnarray}
\noindent 

with $\Re \lbrace..\rbrace$ the real part.

In the above integrand we  define 
\begin{equation}
e^{-i\phi(t)}=e^{-i\varepsilon_0 t-i\delta\phi(t)}\sum_{nm} J_n(x_1)e^{i n(\omega_0 t+
\alpha)}J_m(x_2)e^{-i2 m \omega_0 t} \, ,
\end{equation}
\noindent where we have used the expansion of $\exp ( i x \sin u)=\sum _p J_p(x)\exp (-i p u) $ 
in terms of Bessel functions of first kind.  
We also defined  $x_1=A_1/\omega_0$ and $x_2=A_2/(2\omega_0)$.

For a  gaussian  white noise, the correlator is $\langle \delta\varepsilon(t)\delta\varepsilon(t')\rangle= 2 \Gamma_2
\delta (t - t') $. As   
 $\delta \phi(t)=\int_{0}^t \delta\varepsilon(\tau) d\tau $, 
we take the average over noise in Eq.(\ref{rate}) using
 $\langle e^{i\delta\phi(t)}e^{-i\delta\phi(t')}\rangle=e^{-\Gamma_2|t-t'|}$, with $\Gamma_2$
 the decoherence (pure dephasing) rate  for this model of classical diagonal noise.

The next step is to perform the time integration in Eq.(\ref{rate}),  getting 
\begin{eqnarray}
&&W=\frac{\Delta^2 \Gamma_2 }{2}{\Re}\lbrace\sum_{nn'mm'}\lambda_{nn'mm'}\times \nonumber \\
&& \frac{e^{i(n-n')\alpha}e^{i\omega_0(n-n')t}e^{2i\omega_0(m'-m)t}} {(\varepsilon_0 + (2m -n)\omega_0)^2+\Gamma_2^2} \; \rbrace,
\end{eqnarray}
\noindent

with $\lambda_{nn'mm'} \equiv J_n(x_1)J_{n'}(x_1)J_m(x_2)J_{m'}(x_2)$.

Under the fast driving regime, the non zero exponents  are highly
oscillating compared to the  time scale of the 
driving. Therefore  we only keep
 the contributions with  $\omega_0(n-n')+2\omega_0(m'-m)= 0$, 
and the transition rate reads:
%Alternative, the averaging over one period of the driving gives exactly the
%same result.

\begin{equation}\label{eqb}
W=\frac{\Delta^2\Gamma_2}{2}\sum_{nn'mm'}\frac{\lambda_{nn'mm'} \cos{((n-n')\alpha)}}{(\varepsilon_0 + (2m -n)\omega_0)^2+\Gamma_2^2} \delta_{n-n',2(m-m')} .
\end{equation}

\vspace{0.5cm}
Equation (\ref{eqb}) exhibits an explicit dependence on the phase lag $\alpha$, and a lorentzian line shape close to  the resonance condition  $\varepsilon_0 =(n-2m)\omega_0$, which is characteristic of  white noise models in the regime of times $t\ll T_1$. 
For $x_2= 0$ ,  Eq.(\ref{eqb}) reduces to the expression of the transition rate obtained for single driving protocols.\cite{amin,berns}

\begin{figure}[h]
\begin{center}
\includegraphics[width=1.\linewidth,clip]{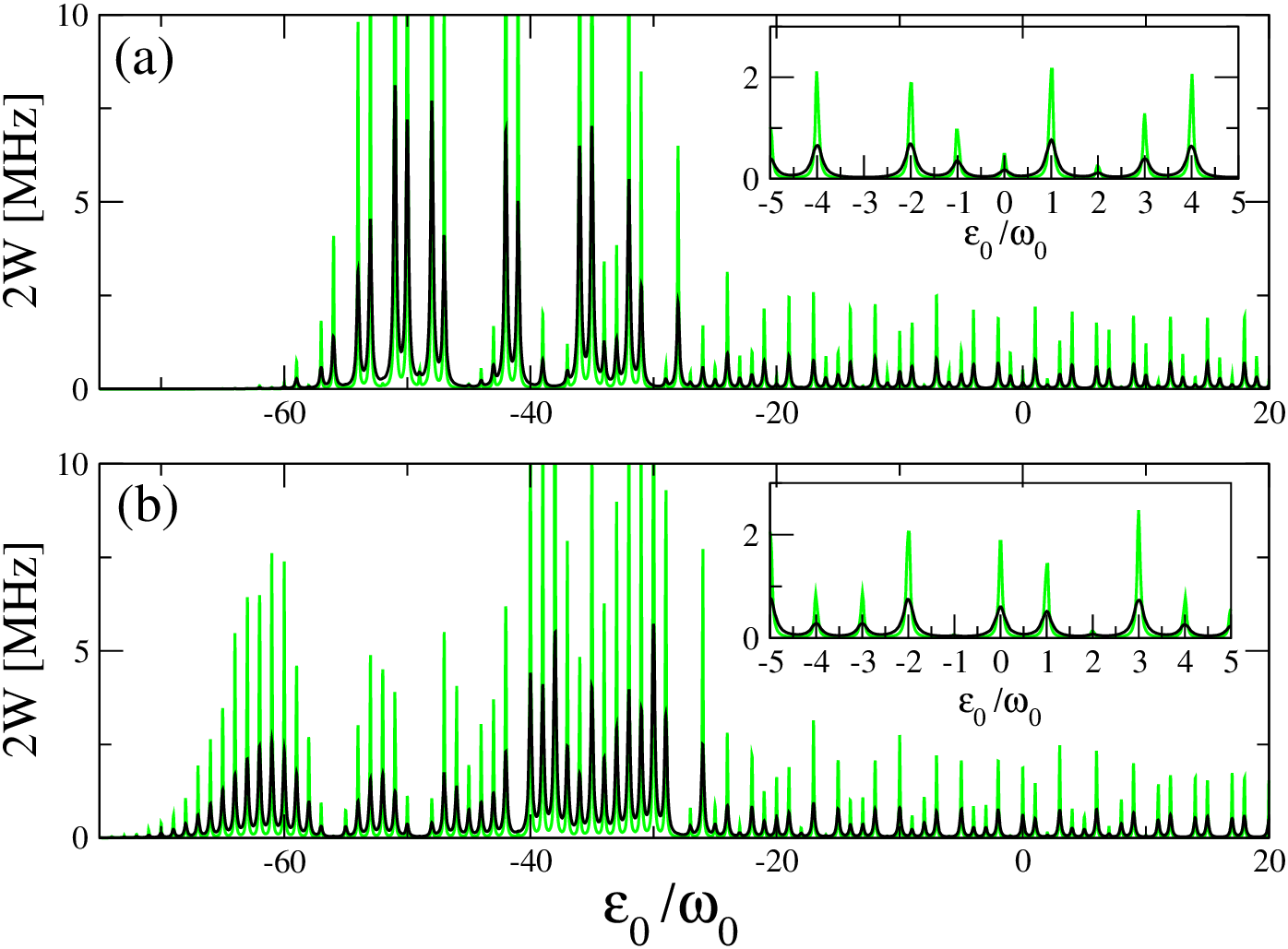}
\end{center}
\caption{(color online)  Rate $2 W $ obtained from 
Eq.(\ref{eqb}) for FQ parameters $\Delta/h=19\,$ MHz, $\omega_0/(2\pi)=
125\,$ MHz, $A_{1}=3\,m\Phi_0$, $A_{2}=1.65\,m\Phi_0$, for $\Gamma_2=100\,$ MHz (black line) and $\Gamma_2=30\,$ MHz (green line). (a) $\alpha=0$ , (b) $\alpha=0.2$.
The insets show a magnification of selected  resonances.}  \label{f2}
\end{figure}

In  Fig.\ref{f2} we  plot  $2 W$ obtained from Eq.(\ref{eqb}) as a function of the static flux detuning $\varepsilon_0$, for the symmetric driving, $\alpha=0$, in  Fig.\ref{f2}(a) and for  $\alpha= 0.2$ in  Fig.\ref{f2}(b). The FQ  parameters are  $\Delta/h=19\,$ MHz, 
$\omega_0/(2\pi)=125\,$ MHz, $A_{1}=3 \,m\Phi_0$ and $A_{2}=1.65\,m\Phi_0$,  identical to those reported in the experiment of Ref.\onlinecite{gustavsson}. 
 The salient feature is that the  peaks are not symmetric with $\varepsilon_0$, exhibiting a  more fluctuating pattern of resonances for $\varepsilon_0 < 0$. This is a manifestation of the sensitivity  of the total  accumulated phase in one period of the driving with $\varepsilon_0$ and $\alpha$.
We have chosen the same amplitudes ratio  $A_{2}/A_{1}= 0.55$ as in  Ref.\onlinecite{gustavsson}, selected to  drive  the FQ  up to four times through the avoided crossing in one driving  period $\tau$, in  the range of  negative 
detunings $\varepsilon_0 < 0$ (see Fig. \ref{f1}).
The strong fluctuating pattern is due to  the three different  phases (one phase for two successive passages) and eight possible superposition states that arises as  $\varepsilon_0$ is varied.  
For other values  of $\varepsilon_0$, the   waveform traverses the avoided crossing zero or   two  times per cycle, producing no accumulated  phase  or a single one, with interference
conditions that  originate a smoother behaviour of $\Gamma$ with $\varepsilon_0 $.

As expected, the resonance peaks  decrease and  turn  wider  as the dephasing rate $\Gamma_2$, included in Eq.(\ref{eqb}) as a parameter, is increased. 
This  is fully consistent with the transition from the non overlapping to the overlapping resonances limit, also observed experimentally  for the single harmonic driving protocols. \cite{oliver,valenzuela}

The derivation can be adapted  to consider  other spectral functions with Gaussian noise. In 
the case of FQ, the magnetic flux noise in SQUIDS could have  a spectral density $S(\omega)$, which for low frequencies behaves as $ 1/ \omega ^ p$ noise. \cite{paladino,amin}
For this case, we get for  the transition  rate: 

\begin{eqnarray}\label{1over}
W_{lf} & & = \Delta^2 \sqrt {\frac{\pi}{ 8 \Gamma_2}}\sum_{nn'mm'}\lambda_{nn'mm'} \cos{((n-n')\alpha)}\times \\ \nonumber 
& & {\exp \lbrace\frac{- (\varepsilon_0 + (2m - n)\omega_0)^2}{\Gamma_2}\rbrace}
 \; .
\end{eqnarray}
where  $\Gamma_2= \int S(\omega) d\omega$ is assumed finite, and we define $W_{lf}$ as the transition rate induced by driven in the presence of low frequency noise.
Notice that the main effect of a noise with a low frequency part is to modify the lorentzian line shape of the individual resonances by a Gaussian line shape.

In  Ref.\onlinecite{gustavsson}, the  transition rate induced by driving  $W$  was  fully identified  with the equilibration rate $\Gamma$  (up to a factor 2). As we  anticipated,  this requires  
 of values   $W < \Gamma_2$,  and  time scales $t \ll T_{1} = 1/\Gamma_1$. 
 
We will show in the next section that  new characteristics  emerge in the behaviour
of the equilibration rate  when the full dynamics, including quantum noise, is considered within the Floquet Markov approach.
In particular, we will analyse the  behavior and sensitivity of the decoherence and relaxation rates with the flux detuning. The  strong variations that these two rates experience close to   resonances with the driving field, question the results of this section  for times scales close to full relaxation.

\subsection{Equilibration rate within the Floquet Markov Master Equation}
\label{fm}

We start this section by reviewing the main ingredients of the
Floquet Markov formalism, and for  further details we suggest  
the Appendix of Ref.\onlinecite{ferron_arx}.

Since the FQ is driven with a biharmonic magnetic flux 
$f(t)= A_{1}\cos{(\omega_0 t+\alpha)}-A_{2}\cos{(2\omega_0 t)}$, its hamiltonian 
is time periodic ${\cal H}(t) = {\cal H}(t + \tau)$, with 
$\tau=2\pi/\omega_{0}$. 
In the  Floquet formalism, which allows to treat  periodic forces
of arbitrary strength and frequency \cite{floquet,fds,ferron_prl,ferron_arx},
the solutions of the    Schr\"odinger equation are of the
form  $|\Psi_\beta(t)\rangle=e^{i\mu_\beta t/\hbar}|\beta(t)\rangle$, where
the  Floquet states $|\beta(t)\rangle$
satisfy $|\beta(t)\rangle$=$|\beta(t+ \tau)\rangle = 
\sum_k |\beta_{k} \rangle e^{-ik\omega_0 t}$, and
are eigenstates of the equation
$[{\cal H} (t)- i \hbar \partial/\partial t ] |\beta(t)\rangle= \mu_\beta |\beta(t)\rangle$,
with $\mu_\beta$ the associated quasi-energy.

Experimentally, the FQ is affected by the electromagnetic environment that  introduces 
decoherence and relaxation  processes.
A  standard theoretical model to cope with these effects, is to linearly couple
the system to a bath of harmonic oscillators 
with a spectral density
$J(\omega)$ and   equilibrated at  temperature $T$.\cite{hanggi,milenas,grifonih,caspar}
For the FQ the dominating source of noise is flux-like,
in which case the bath degrees of freedom couple with
the system variable $\varphi_l$ \cite{caspar, ferron_arx}
In the two-level representation of  Eq.(\ref{htls}), the flux noise is usually represented
by a system bath hamiltonian of the form $H_{sb} \propto \sigma_z$. \cite{milenas}

For  weak coupling  (Born approximation) 
and fast bath relaxation (Markov approximation),
a Floquet-Born-Markov master  equation  
for   the reduced density matrix $\hat\rho$
in the Floquet basis,
$\rho_{\alpha\beta}(t)=\langle\alpha(t)|\hat\rho(t)|\beta(t)\rangle$,
can be obtained.\cite{hanggi}
 
Considering that the time scale $t_r$ for full relaxation satisfied
$t_r \gg \tau$, one gets (see  Appendix of Ref.\onlinecite{ferron_arx} for details):
\begin{equation}
\label{drho}
\frac{d\rho_{\alpha\beta}(t)}{dt}=-\frac{i}{\hbar}
(\mu_\alpha-\mu_\beta)\rho_{\alpha\beta}(t) +%\nonumber\\
\sum_{\alpha'\beta'}{\it L}_{\alpha'\beta'\alpha\beta} \; \rho
_{\alpha'\beta'}(t) \; .
%\label{rw1}
\end{equation}

The first term in Eq.(\ref{drho}) represents the non dissipative 
dynamics and the influence of the bath is described by the
 rate coefficients averaged over one period of the driving $\tau$,  \cite{ferron_arx}
%\begin{equation}
%{\cal L}_{\alpha\beta\alpha'\beta'}(t)= \sum_q{\cal L}^{q}_{\alpha\beta\alpha'\beta'}e^{-iq\omega_{0} %t}
%\end{equation}
\begin{eqnarray}
\label{superla}
{\it L}_{\alpha\beta\alpha'\beta'}&=& R_{\alpha\beta\alpha'\beta'}
+R_{\beta\alpha\beta'\alpha'}^* \\
& &-\sum_\eta \left( \delta_{\beta \beta'}
R_{\eta\eta\alpha'\alpha}+ 
\delta_{\alpha \alpha'}R_{\eta\eta\beta'\beta}^* \right).
\nonumber
\end{eqnarray} 
The rates 
\begin{equation}
R_{\alpha\beta\alpha'\beta'} = \sum_{q} g_{\alpha \alpha'}^qA_{\alpha \alpha'}^qA_{\beta' \beta}^{-q} \; ,
\label{rates}
\end{equation}
can be interpreted as  sums of $q$-photon exchange terms and contains  information  on the system-bath coupling operator $A_{\alpha\beta}^q=\sum_{nm}\sum_{k} \alpha_{k,n}^*\beta_{k+q,m}\langle n|\varphi_l|m\rangle$ written down in  
terms of the eigenbasis $|n\rangle$ of the  FQ Hamiltonian for the undriven case, Eq.(\ref{ham-sys}), with  $\alpha_{k,n}= \langle n|\alpha_k\rangle$.  
The nature of the bath is encoded  in the 
coefficients
 $g_{\alpha \beta}^q=J(\mu_{\alpha\beta}^q/{\hbar})n_{\rm th}(\mu_{\alpha\beta}^q)$ 
with $\mu_{\alpha\beta}^q=\mu_\alpha-\mu_{\beta}+q\hbar \omega_{0}$
and $n_{\rm th}(x)=1/(\exp{(x/k_BT)}-1)$. 
Here we consider the FQ in contact with an Ohmic bath with a spectral density $J(\omega)=\gamma \omega$ (with a cutoff frequency),
defining $J(-x)=-J(x)$ for $x<0$, but other spectral densities could be included.\cite{ferron_arx} 

The Floquet Markov formalism   has been extensively
employed to study relaxation and decoherence 
in double-well potentials and two level systems driven by  single frequency periodic evolutions . \cite{hanggi,grifoni-hanggi,breuer,ketzmerick,grifonih} More recently we applied it to analyze the  FQ under strong harmonic driving \cite{ferron_prl, ferron_arx}, 
when many energy levels have to be taken into account.

 As in the  present work  the dynamics of the FQ  under a weak  biharmonic driving protocol  is studied, in the following the FQ Hamiltonian  will be reduced to its lowest two levels, Eq.~(\ref{htls}).
%  the treatment of the full multilevel Hamiltonian of the FQ 
% Eq.~(\ref{ham-sys}) to Sec.\ref{ml}. \cite{ferron_prl}
%Now, we can calculate the coefficients ${\it L}_{\alpha'\beta'\alpha\beta}$
%using the Floquet states $|\Phi_\beta^{k} \rangle$ and quasienergies 
%$\varepsilon_\beta$, obtained
%as in \cite{fds}, and then we integrate numerically Eq.~(\ref{drho}), 
%obtaining $\rho_{\alpha\beta}(t)$ as a function of $t$.

For large times scales, it is usually assumed that
the density matrix becomes approximately diagonal in the
Floquet basis \cite{grifonih}. This approximation
is justified when  $\mu_\alpha - \mu_\beta \gg {\it L}_{\alpha'\beta'\alpha\beta} $, which is fulfilled for very weak coupling with the environment and away from
resonances.\cite{breuer,ketzmerick,fazio1,ferron_arx} 
However, to  compute fluctuation effects  it is necessary to  sweep in  the dc flux detuning $\epsilon_0$, attaining near resonances quasi degeneracies,  {\it i.e.}  $\mu_\alpha - \mu_\beta \sim 0$. Therefore as the dynamics of the diagonal and  off-diagonal density matrix can not be separated, we have to solve the full Floquet Markov equation, Eq.(\ref{drho}), to find  relaxation and decoherence rates close to resonances .

The rates are extracted from the non zero  eigenvalues  of the matrix $\hat{\Lambda}$, given from its entries in the Floquet basis, ${\it L}_{\alpha\beta\alpha'\beta'}$, defined in Eq.(\ref{superla}).
The long time relaxation rate $\Gamma_r = 1/t_r$  is the minimum real eigenvalue
 (excluding the eigenvalue $ 0$). In addition, the
decoherence    rate  $\Gamma_{ab}$  is given by the negative real part of the
complex conjugate pairs of eigenvalues of $\hat{\Lambda}$.\cite{ferron_arx}

%The off-diagonal part is dominated by the dependence
%$$ \frac{d\rho_{\alpha\beta}}{dt} \approx \left[-\frac{i}{\hbar}
%(\varepsilon_\alpha-\varepsilon_\beta)+ L_{\alpha\beta\alpha\beta}\right] \,\rho_{\alpha\beta}\;\;\;\; \alpha\not=\beta$$
%In this approximation, the decoherence rate between the $|\alpha\rangle$ and
%the $|\beta\rangle$ Floquet state, 
%is given by $\Gamma_{\alpha\beta}=-L_{\alpha\beta\alpha\beta}$.
%The dynamics for the diagonal part of the density matrix gives
%a rate equation for the population of the Floquet states
%$P_\alpha=\rho_{\alpha\alpha}$:
%\begin{eqnarray}
%\frac{dP_\alpha}{dt}&=&\sum_{\beta} L_{\alpha\alpha\beta\beta} P_\beta\nonumber\\
%&=& 2\sum_{\beta}R_{\alpha\alpha\beta\beta}P_\beta - R_{\beta\beta\alpha\alpha}P_\alpha
%\end{eqnarray}
%where $R_{\alpha\alpha\beta\beta}=\sum_n
%g_{\alpha\beta}^n|A_{\alpha\beta}^n|^2$, after Eq.~(\ref{rates}).

The sensitivity of  both   rates, $\Gamma_{ab}$ and   $\Gamma_r$, on the dc flux detuning and  the driving amplitude  has been analyzed  in recent  studies of the phonomena of population inversion and dynamic transitions in the LZS interference patters of (single) driven FQ. Both  phenomena  emerge away from resonances, in the long time  regime. \cite{ferron_prl,ferron_arx}
 
In addition, as we show below, close to resonances decoherence and relaxation rates will  attain  values  much larger than those predicted for the undriven case.
 
\begin{figure}[h]
\begin{center}
\includegraphics[width=.8\linewidth,clip]{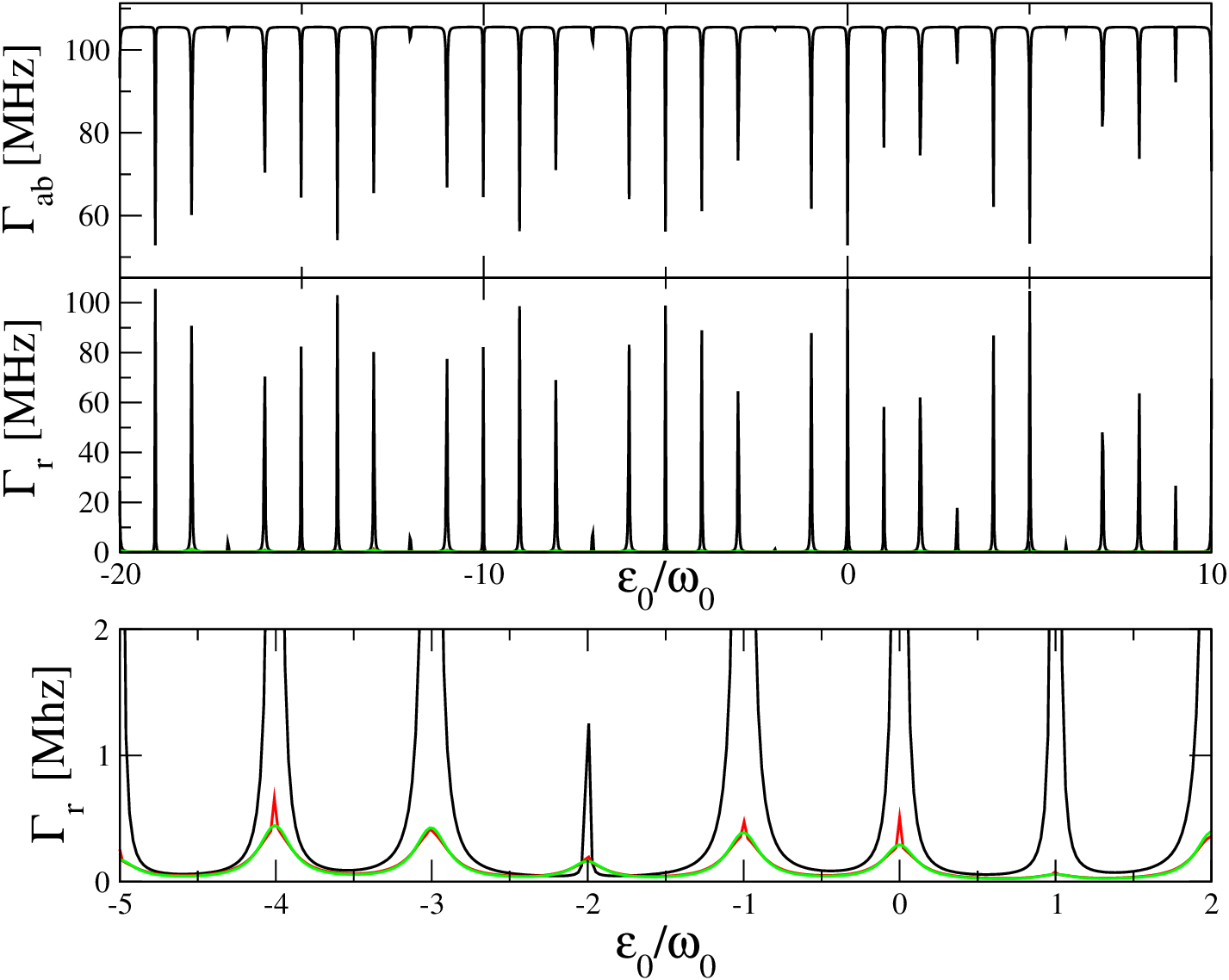}
\end{center}
\caption{(color online) Upper panel: decoherence rate $\Gamma_{ab}$ obtained in the FM formalism,  as a function of the normalized flux  detuning $\epsilon_0/\omega_0$.
Intermediate panel:  relaxation rate $\Gamma_r$ as a function of  (normalized)  flux detuning obtained within the FM formalism (black solid line). Rate $2 W$ obtained  from  Eq.(\ref{eqb}) with  $\Gamma_2=100$ MHz (green line) and after replacing   $ \Gamma_2 \rightarrow \Gamma_{ab}$ (red line). The lower panel shows an enlarge view  to stress the differences between $\Gamma_r$ and $2 W$ 
close to resonances. Parameters are $\Delta/h=19\,$ MHz, 
$\omega_0/(2\pi)=125\,$ MHz, $A_{1}=3 \,m\Phi_0$ and $A_{2}=1.65\,m\Phi_0$. For the FM calculations
the bath is ohmic, $J(\omega)= \gamma \omega$, at temperature $T= 20mK$  with coupling $\gamma= 0.001$. }  \label{f3}
\end{figure}

In Fig.\ref{f3}, $\Gamma_{ab}$ and $\Gamma_r $ are plotted as a function of the normalized flux detuning $\epsilon_0/\omega_0$ - to visualise the resonances positions at integer values.
The calculations  were performed for the same FQ parameters  and driving protocol as in Fig.\ref{f2}, and for an ohmic bath at T= 20 mK, which is the temperature reported  in the experiment of Ref. \onlinecite{gustavsson}.

Both rates exhibit a strong dependence with the detuning and important variations at resonances.
In the case of the decoherence rate
$\Gamma_{ab}$ (Fig.\ref{f3} upper panel), although its value away from resonances is of the order of  $\Gamma_{ab} \sim 100 $ MHz - similar to the decoherence rate  $ 1/ T_{2} =100$ MHz in  Ref.\onlinecite{gustavsson}; the important variations displayed at resonances redound in effectively doubling the reported decoherence time. 

The rate $\Gamma_r$  is  plotted in the lower panels of Fig.\ref{f3} together with  $2W$, computed from   Eq.(\ref{eqb}) for  a  constant value of 
the parameter $\Gamma_2= 100 $ MHz (green line), and  for  $\Gamma_2$ replaced by $\Gamma_{ab}$  obtained within the FM formalism (red line), to include the  dependence with the dc flux already described.  The nominal values of $2W $ and  $\Gamma_r$ away from resonances are quite similar and  even the positions of the resonances are well captured in both cases (see the lower panel for a blow up).
However, at resonances  $\Gamma_r$ can attain values close to $100 $ MHz, satisfying  $\Gamma_r\sim 2 \Gamma_{ab} $.
This  is expected  for a longitudinal system- bath coupling (as the one considered in the present work, {\it i.e.} $H_{sb} \propto \sigma_z$)  and is a fingerprint of  the suppression of a pure dephasing mechanism on resonance condition. \cite{grifonih,ferron_arx}
Notice that when $\Gamma_r \sim 2 \Gamma_{ab}$ the time scale separation  $T_1 \gg T_2$- which was assumed in the experiment of Ref.\onlinecite{gustavsson} and in the phenomenological approach developed in Sec.\ref{stls}- is not  fulfilled.

\begin{figure}[h]
\begin{center}
\includegraphics[width=1.\linewidth,clip]{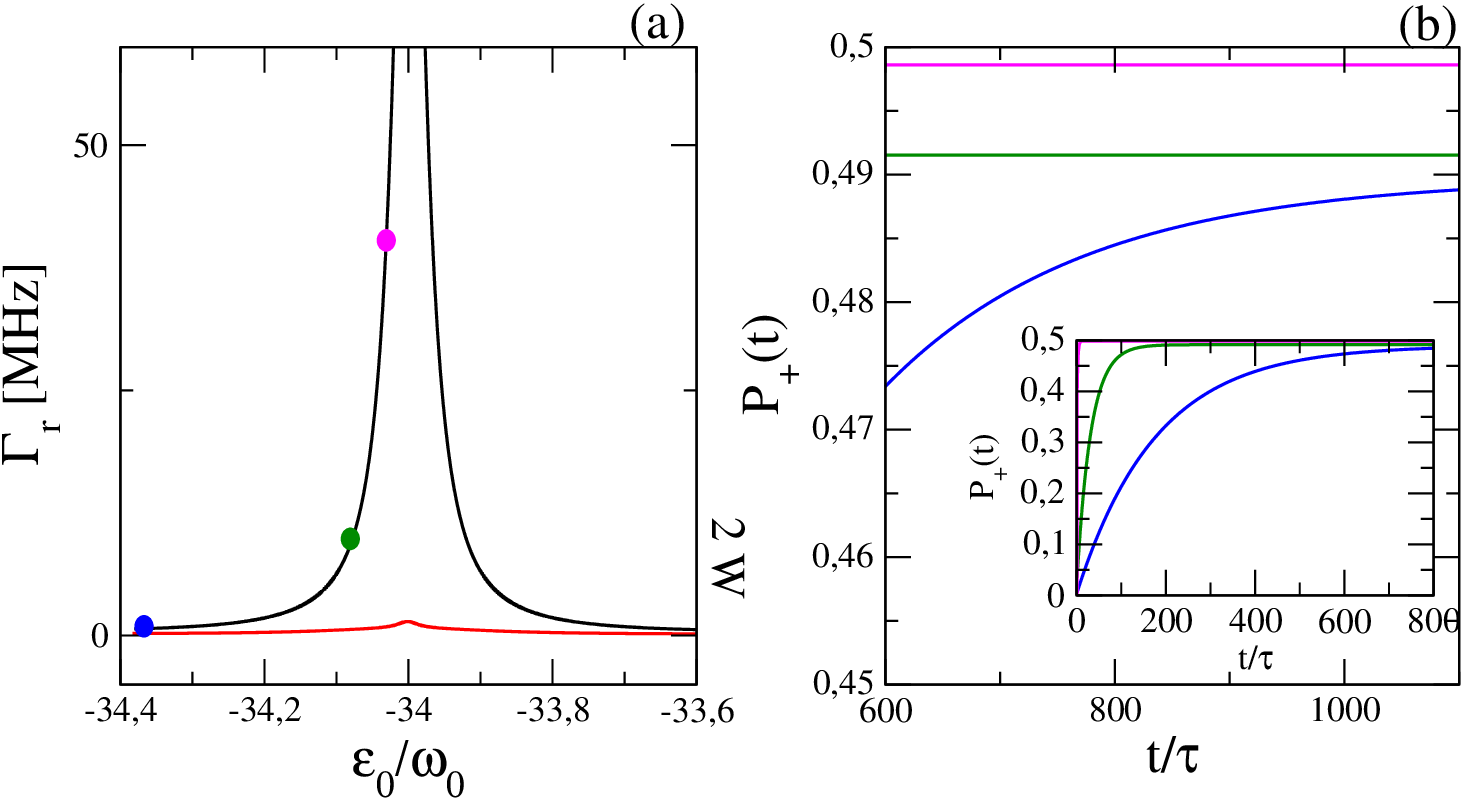}
\end{center}
\caption{(color online) (a)  Relaxation rates  rate $\Gamma_{r}$ close to a resonance obtained within the FM formalism   as a function of the normalized flux  detuning $\epsilon_0/\omega_0$ and for the same parameters than in Fig.(\ref{f3}).
The red curve corresponds to $\Gamma= 2 W$ obtained  from Eq.(\ref{eqb}), after replacing   
$ \Gamma_2 \rightarrow \Gamma_{ab}$. The chosen resonance corresponds to $\epsilon_0/\omega_0= -34 $.
(b) Excited state ocupation probability $P_{+} (t)$ as a function of the normalized time $t / \tau$, obtained for  three sampled values of $\Gamma_{r}$ (identified by the dots in panel (a)). In each case,  the initial state is the ground state of the TLS Hamiltonian Eq.(\ref{htls}) for the correspondent flux detuning $\epsilon_0/\omega_0$ }  \label{f4}
\end{figure}

Although the resonance condition could seem  very sharp to be  experimentally fulfilled, significant increments in the values of $\Gamma_r$ relative to the background values can be also appreciated in a close vicinity, as it is displayed in  Fig.\ref{f4}(a).
%As we show in fig.\ref{f4}(a), even away from
%the resonant condition, the values of $\Gamma_r$ obtained within the FM approach are still
%much larger than those of $\Gamma= 2 W$.

Relative changes of  $\sim 10-20 \%$ in  $\epsilon_{0}/\omega_0$,  produce  concomitant variations in  the values of  $\Gamma_r$ which give rise to different  profiles for  the associated excited state  occupation probabilities $P_{+} (t)$ (see  Fig.\ref{f4}(b)).
 Even  for time scales 
$t\sim t_{exp} \sim 1000 \tau $   appreciable differences still persist in the respective $P_{+} (t_{exp})$. 

\section{Computing  averages and fluctuations in the  rates: The  mesoscopic analogy}
\label{fluct}

The total accumulated phase during the driving protocol is controlled by the asymmetry parameter  $\alpha$, which modifies the  biharmonic waveform. As a consequence,  $2 W$  and  $\Gamma_r$ should experience, besides  the sensitivity  on $\epsilon_0$,   fluctuations as a function of $\alpha$. \cite{gustavsson}

As we already mentioned, the transition rate induced by the driving, $W$, was  identified
in  Ref.\onlinecite{gustavsson}   with 
an effective transition amplitude -the  essential ingredient to determine
the conductance in  the Landauer formalism. \cite{datta}
Under the assumption   that the equilibration rate can be written as $\Gamma = 2 W$, the approach  was to associate the  fluctuations in $\Gamma$  as function of  the dc flux, with the  Universal conductance fluctuations (UCF), in analogy with mesoscopic electronic systems.\cite{varios_ucf}

In the previous section, we have seen that  for time scales close to
full relaxation and around  resonances with the driving field, important and quantitative differences emerge between $2 W$ and the relaxation (equilibration) rate, $\Gamma_r$, obtained within the FM formalism. Associated with this, the respective fluctuation patterns will also exhibit different behaviours, as we  show in the following.
 
The averages $<2W>$  and   $<\Gamma_r>$, over  $\varepsilon_0$, are
defined as $
\langle ...\rangle=\frac{1}{\varepsilon^{max}}\int_{0}^{\varepsilon^{max}}...\,d\varepsilon_0 \, 
$, and play the role of   ensemble averages over different  scatterers configurations.
In the case of the phenomenological approach we  computed from  Eq.(\ref{eqb}):
\begin{eqnarray}
&&\langle W\rangle=\frac{\Delta^2}{2\varepsilon^{max}} \sum_{nn'mm'}\lambda_{nn'mm'} \cos{((n-n')\alpha)} \label{av} \\
&&\left[\arctan{\left(\frac{(2m - n)\omega_0}{\Gamma_2}\right)}
-\arctan{\left(\frac{(2m -n )\omega_0-\varepsilon^{max}}{\Gamma_2}\right)}\right] \nonumber  \;,
\end{eqnarray}

and

\begin{equation}\label{av2}
\langle W^2 \rangle= \frac{1}{\varepsilon^{max}}\int_{0}^{\varepsilon^{max}} W^2  d\varepsilon_0 \;.
\end{equation}

Although  $\langle W^2\rangle$ does not have a simple analytic expression, 
 its numerical evaluation is straightforward. The fluctuations in
 $2W$ are  defined as
$\sigma_{2W}=2\sqrt{\langle W^2\rangle-\langle W\rangle ^2}$
%with $\langle W^2\rangle=\frac{1}{\varepsilon^{max}}\int_{0}^{\varepsilon^{max}}W^2 $.

In the case of the relaxation rate $\Gamma_r$, its  average $\langle\Gamma_r \rangle$  and fluctuations  $\sigma_r$, have been computed numerically.

\begin{figure}[h]
\begin{center}
\includegraphics[width=1.\linewidth,clip]{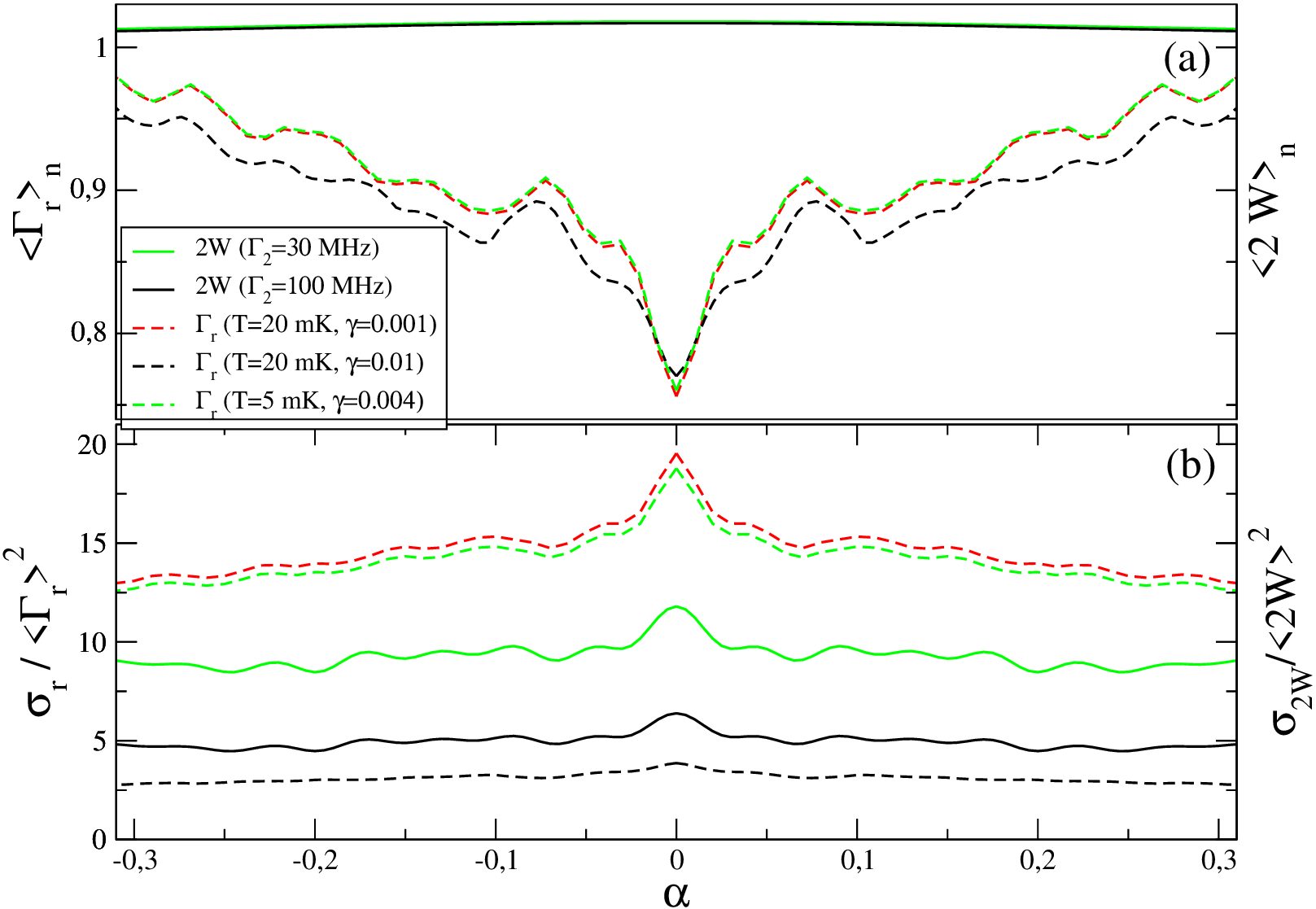}
\end{center}
\caption{(color online) normalized rates $<\Gamma_{r}>_n$ and $<2W>_n$ (panel (a)) and
normalized standard deviations $\sigma_r /<\Gamma_r>^2$ and  $\sigma_{2W}/< 2W>^2$ (panel (b))
 averaged from $-4m\Phi_0$ to $0m\Phi_0$   as a function of the time reversal broken parameter $\alpha$.
The different values of the dephasing rate $\Gamma_2$ used to compute $\Gamma= 2W$ and the values of the temperature $T$ and bath coupling constant $\gamma$ employed to
compute the  rates $\Gamma_r$  within the FM formalism,  are respectively specified in the inset.
FQ parameters are $\Delta/h=19\,$ MHz, $\omega_0/(2\pi)=125\,$ MHz, 
$A_{1}=3 \,m\Phi_0$ and $A_{2}=1.65\,m\Phi_0$.} \label{f5}
\end{figure}

In Fig.\ref{f5}(a) we plot the averaged  rates relative  to  its values at $\alpha=0.5$, {\it i.e} 
$\langle \Gamma_{r} \rangle_{n} \equiv\langle \Gamma_r \rangle/\langle \Gamma_r\rangle_{(\alpha=0.5)}$  and $\langle 2W\rangle_{n}\equiv\langle 2W\rangle/ \langle 2W\rangle_{(\alpha=0.5)}$, in order to establish a fair comparison for different bath temperatures, couplings $\gamma$ or dephasing rates $\Gamma_2$, respectively.
  
Even after performing the averages in flux detuning, strong fluctuations are still visible  as a function of  $\alpha$.
Notice that  $\langle 2 W\rangle_n$ is almost independent of $\alpha$ and the dephasing rate $\Gamma_2$, in agreement with the results of Ref.\onlinecite{gustavsson}. On the other hand, $\langle \Gamma_{r} \rangle_{n}$ exhibit a sharp dip at $\alpha=0$, that  could  be interpreted as the fingerprint of the Weak Localization (WL) correction- in analogy to the correction present in the
quantum conductance of  mesoscopic disordered systems. \cite{varios_wl} 
The relative fluctuations (normalized to the squared mean values) are defined as
$\sigma_r /\langle \Gamma_r\rangle^2 $ and $\sigma_{2W}/\langle 2W\rangle^2$,  and are plotted  in Fig.\ref{f5}(b). The profiles resembles the UCF found in short mesoscopic wires, with the fluctuations at $\alpha=0$ enhanced compared to the fluctuations for $\alpha\neq 0$, as the theory of UCF predicts when the time revesal symmetry is broken. \cite{varios_ucf}

The WL correction  and the UCF  tend to  wash   out as  either the effective temperature and/or the
coupling with the environment are increased, as expected when decoherence and relaxation processes
become more efficient. This is clearly observed in Figs.\ref{f5}(a) and (b).
In addition, the  profiles remain almost undisturbed when the product of the effective temperature $T$ times the bath coupling $\gamma$ remains constant, although each one is varied separately.
This is consistent with the well known result that the dominant contribution to the  decoherence rate depends on the product $\gamma T$. \cite{caspar}

We want to point out some limitations of the phenomenological approach employed in Sec.\ref{stls} and followed in Ref.\onlinecite{gustavsson}.
On one hand, it  disregards relaxation assuming that the equilibration rate $\Gamma$ is given by $ 2 W$. As a consequence the equilibration rates result  largely underestimated close to resonances,  as we emphasized when describing Fig.\ref{f3}.
On the other hand, the interpretation of  $\Gamma$ as a conductance is only
well justified away from resonances, when is satisfied that $\Gamma \sim 2 W$.

Last but not least, we want to comment on the detection of the WL effect, which has not been measured in  Ref.\onlinecite{gustavsson}. 
To our view the limitation which precludes the experimental  observation of the  WL-like effect is not the driving protocol, as the authors of Ref.\onlinecite{gustavsson} suggested,
but mainly the difficulty in capturing the extremely sharp resonant conditions as the flux detuning is swept, and also the relatively short experimental decoherence times $T_2$.
From the theory of disordered electronic systems  it is known that
the size of the Weak Localization correction scales (logarithmically) with  the dephasing time $\tau_{\phi} \propto T_{2}$.\cite{varios_wl}

%value of the  critical magnetic field needed  to destroy the WL correction scales as  $B_{c} \propto %1/ \tau_{\phi}$, with $\tau_{\phi} \propto T_{2}$, the dephasing time. 
%Applying  this analysis  to the present case, and  defining  the critical value  to suppress the WL-like correction as $\alpha_c$, it should scale  as $\alpha_c  \propto 1/T_{2} = \Gamma_2$.
Consistent with this result, Fig.\ref{f5}(a) shows very well defined dips in  $\langle\Gamma_r \rangle_n$ at $\alpha =0$, for values of the coupling
$\gamma = 0.004\sim 0.001$ and  temperature of 20 mK-as the reported experimentally.
These values  give decoherence  rates $\Gamma_2\sim  25-30 $ MHz ($T_2= 1/\Gamma_2 \simeq 30-40$ns). 
Thus it is expected that for  slightly larger values of $T_2$- but not so far from the  reported    $T_2\sim 10 ns$, it could be possible  to experimentally access to the full WL dip, once the resonance conditions can be explored.

%\begin{figure}[h]
%\begin{center}
%\includegraphics[width=0.8\linewidth,clip]{fig6.eps}
%\end{center}
%\caption{(color online) rate $\Gamma= 2W$ averaged from $-4m\Phi_0$ to 
%$0m\Phi_0$ and Standard deviation for qubit 
%parameters $\Delta/h=19\,MHz$, $\omega_0/(2\pi)=125\,MHz$, 
%$A_{1}=3 \,m\Phi_0$ and $A_{2}=1.65\,m\Phi_0$
%computed using Eqs.(\ref{eqb}), (\ref{av}) and (\ref{av2}) for decoherence rates $\Gamma_2=100\,MHz$ %(red line), and $\Gamma_2=30\,MHz$ (blue line).
%(a)Standard Deviation of the transition rate $\sigma$ and (b)  $<\Gamma>$.} \label{f6}
%\end{figure}

\section{Conclusions and perspectives}
\label{conc}

In this work we have tested fluctuation effects associated to  broken time reversal symmetry in FQ 
driven by a biharmonic (dc + ac) magnetic flux with a phase lag.

Employing the full Floquet Markov Master Equation we have computed relaxation $\Gamma_{r}$ and decoherence $\Gamma_{ab}$ rates, both resulting strongly dependent on  the phase lag and the dc flux detuning, exhibiting appreciable fluctuations.
As a function of the dc flux and away from the resonance conditions with the driving field, $\Gamma_{ab} \rightarrow \Gamma_{2} $ and  $\Gamma_{r} \rightarrow \Gamma_{1}$ with  $\Gamma_{2}=1 /T_{2} >>\Gamma_{1}= 1 /T_{1}$, {\it i.e.} in agreement with the relaxation and decoherence rates predicted for the undriven FQ.
However, close  to  resonances both rates show large variations compared to the respective values out of resonances, even satisfying  $2 \Gamma_{ab} \sim \Gamma_{r}$.

The relaxation (equilibration) rate $\Gamma_{r}$ can be  accessed experimentally by measuring the  decay of the FQ excited state population. Recently the  fluctuations in the measured equilibration rate have been analized and  associated  to   Universal conductance fluctuations (UCF), following  an  analogy to well known phenomena exhibited in disordered mesoscopic electronic systems. \cite{gustavsson} However, as we discuss in extent along this work, the {\it mesoscopic  analogy} is only well  justified for the out of resonance regime, when the equilibration rate can be described in terms of a transition probability induced by the driving protocol.

Besides UCF, we also predict that the Weak Localization effect can be detected
 for the biharmonic driving protocol. However to observe this effect, the experiments should be performed in a more coherent 
regime, in  which larger values of $T_2$ could be attained.
Nowadays    the control on the environmental  bath degrees of freedom  is  a promising way 
 to  enlarge coherence, as  have  been  recently proposed and tested. \cite{ferron_prl,paavola,chen}
 
By increasing the driving amplitude, more avoided crossings of the FQ can be reached, 
 enabling  the computation of averages and fluctuations beyond the weak scattering limit- which in fact is a limitation of the present approach to the mesoscopic analogy.
This regime is experimentally attainable as the amplitude spectroscopy experiments \cite{bylan2009} have  proven.

%\cite{fds_ip}  

We acknowledge support from CNEA, UNCuyo (P 06/ C455), CONICET PIP11220150100218, PIP11220150100327 and ANPCyT PICT2011-1537, PICT2014-1382)
%\section*{References}

\end{document}